\documentclass[aps,pra,preprint,superscriptaddress,raggedbottom]{revtex4-1}
\usepackage{amsfonts}
\usepackage{amsmath}
\usepackage{txfonts}
\usepackage{amssymb}
\usepackage{amsbsy} 
\usepackage{epsfig}
\usepackage{cancel}
\usepackage{float}
\usepackage{romannum}
\usepackage{color}

\def\be{\begin{equation}} \def\ee{\end{equation}}
\def\bea{\begin{eqnarray}} \def\eea{\end{eqnarray}}

\def\bB{{\bf B}}

\def\bG{{\bf{G}}}
\def\bn{\hat{\bn}}

\def \bB{{\bf B}}

\begin{document}

\title{Constructing three-dimensional photonic topological insulator using two-dimensional ring resonator lattice with a synthetic frequency dimension}

\author{Qian Lin}
\affiliation{Department of Applied Physics, Stanford University, California 94305, USA}
\author{Xiao-Qi Sun}
\affiliation{Department of Physics, Stanford University, California 94305, USA}
\author{Meng Xiao}
\affiliation{Ginzton Laboratory, Stanford University, California 94305, USA}
\author{Shou-Cheng Zhang}
\affiliation{Department of Physics, Stanford University, California 94305, USA}
\author{Shanhui Fan}
\email[Corresponding email: ]{shanhui@stanford.edu}
\affiliation{Ginzton Laboratory, Stanford University, California 94305, USA}
\affiliation{Department of Electrical Engineering, Stanford University, California 94305, USA}
\date{\today}
\begin{abstract}
In the development of topological photonics, achieving three dimensional topological insulators is of significant interest since it enables the exploration of new topological physics with photons, and promises novel photonic devices that are robust against disorders in three dimensions. Previous theoretical proposals towards three dimensional topological insulators utilize complex geometries that are challenging to implement. Here, based on the concept of synthetic dimension, we show that a two-dimensional array of ring resonators, which was previously demonstrated to exhibit a two-dimensional topological insulator phase, in fact automatically becomes a three-dimensional topological insulator, when the frequency dimension is taken into account. Moreover, by modulating a few of the resonators, a screw dislocation along the frequency axis can be created, which provides robust transport of photons along the frequency axis. Demonstrating the physics of screw dislocation in a topological system has been a significant challenge in solid state systems. Our work indicates that the physics of three-dimensional topological insulator can be explored in standard integrated photonics platforms, leading to opportunities for novel devices that control the frequency of light. 
\end{abstract}

\maketitle
The discovery of two-dimensional (2D) topological photonic systems expands the scope of topological materials from fermions to bosons and enables new ways for controlling the propagation of electromagnetic waves \cite{Haldane2008a,Wang2009a,hafezi2011,fang2012b,Rechtsman2013,Khanikaev2013,chong2013}. 
A natural next step is towards the experimental realization of photonic three-dimensional (3D) topological insulators \cite{qi2011,hasan2010}. However, existing proposals \cite{lu2016a,khanikaev2016} to realize 3D photonic topological insulators require sophisticated geometries. Here we propose to explore such 3D topological physics using a 2D array of ring resonators that is directly implementable in standard on-chip integrated photonic platform. A wide variety of topological phases can be created with the use of 2D arrays of ring resonators \cite{hafezi2011,chong2013,chong2014,fang2012b}. Moreover, since a ring supports a discrete set of resonant modes forming a frequency comb \cite{yuan2016,lin2016,minkov2016,ozawa2016a}, one can consider a ring to have a synthetic third dimension 
along the frequency axis. As a result, a 2D array of ring resonators which behaves as a 2D topological insulator near one of the resonant frequencies of a single ring, will naturally become a 3D weak topological insulator when the additional synthetic dimension is considered. The 3D topological order in this system can be revealed by modulating the rings. For example, by modulating a few resonators in the array, a screw dislocation parallel to the frequency axis can be created, which supports topologically-protected one-way transport \cite{ran2009} along the frequency axis and serves as a spectral probe for 3D topological order.

Three dimensional topological insulator phases are characterized by four indices \cite{fu2007}: a scalar $\nu_0$ and a triad $(\nu_1,\nu_2,\nu_3)$ which defines a reciprocal lattice vector $\bG_\nu$. $\nu_0$ classifies 3D topological insulators into two general classes: weak and strong. The strong class has $\nu_0\ne 0$, and gapless surface states exist on any 2D surface of the sample. On the other hand, weak topological insulators have $\nu_0=0$ but $\bG_\nu\ne\mathbf{0}$. Since $\bG_\nu$, which characterized the topology of a weak topological insulator, depends on crystal orientation, the existence of gapless surface states also depends on the orientation of the surface. Originally assumed to be less robust than their strong counterpart, weak 3D topological insulators have been recently shown to possess rich physics and unexpectedly strong protection against disorder \cite{ringel2012}. In theory weak topological insulators can be realized by simply stacking multiple layers, each being in the 2D topological phase. However, in typical electronic or photonic systems, when stacking multiple layers together, it is difficult to control the interlayer coupling so that the 2D topological gap, which is usually very small, can still persist in 3D. In this letter, we show that a 2D microring array with synthetic frequency dimension naturally realizes a weak 3D topological insulator with minimal interlayer coupling, providing a clean and versatile platform to study the physics of weak 3D topological insulators.

\section*{Results}
\subsection*{Constructing the three-dimensional topological insulator}
We start by constructing a microring array that supports 2D topological states. As an example, we shall use a 2D array that supports a quantum spin Hall effect (QSHE) \cite{hafezi2011}, although the results generalize to all ring array configurations supporting photonic topological states \cite{fang2012b,minkov2016,chong2013}. Consider a Hamiltonian on a square lattice
\begin{equation}
H^{2D}=-t\sum_{x,y,\sigma}\left(a^\dagger_{x+1,y,\sigma}a_{x,y,\sigma}+e^{i\sigma x\phi}a^\dagger_{x,y+1,\sigma}a_{x,y,\sigma}+h.c.\right)
\label{eq:QSHE}
\end{equation}
where $t$ is the strength of nearest-neighbor coupling, $(x,y)$ labels the lattice sites, and $\sigma=\pm1$ represent two pseudo-spins. In this Hamiltonian, the two pseudo-spins are decoupled. A particle with $\sigma=+1$ spin sees a directional phase of $+x\phi$ ($-x\phi$) when hopping up (down) along $y$, corresponding to a uniform magnetic flux of $\phi$ in each square plaquette. A particle with $\sigma=-1$ spin sees the opposite directional coupling phase, and thus a uniform magnetic flux of $-\phi$ per unit cell. This Hamiltonian therefore describes a boson subject to a pseudo-spin-dependent out-of-plane uniform magnetic field, and exhibits QSHE.

The photonic structure that implements this Hamiltonian consists of a square lattice of identical site rings coupled to their nearest neighbors through identical link rings \cite{hafezi2011}, as shown in Fig.~1b. Here the $(x,y)$ labels in Eq.~\ref{eq:QSHE} are pairs of half integers representing the centers of site rings, with the origin of the coordinate system defined at the center of the array. The circumference of the rings are chosen such that the resonance of the site ring matches the anti-resonance of the link ring, and photons are strongly confined in the site rings. The resulting photonic band structure near the resonance of the site ring can be well captured by a tight binding model. In the absence of backscattering in the waveguide forming the ring, the clockwise and counterclockwise propagating modes in the site rings, representing the two spins in QSHE, are degenerate and decoupled. The centers of the $y$-coupling link rings are shifted away from the line segments that connect the center of the nearest neighbor site rings. As a result the left and right branches of the link ring differ in length, and when a photon in the clockwise mode of the site ring hops in the negative $y$-direction it acquires a smaller phase than that when it hops in the positive $y$-direction. The shifts of the link ring vary from one column of the array to the next to provide the $\pm x\phi$ directional coupling phase in the QSHE Hamiltonian in Eq.\ref{eq:QSHE}. We confirm the existence of the 2D topological bandgap in such a microring array by computing the topological edge states on a stripe of the array that is finite in $x$ and infinite in $y$. The magnitude of the shifts are chosen to produce a uniform magnetic field of $1/3$ flux quanta through each square plaquette. We model our system using the transfer matrix method \cite{chong2013} (see Supplementary Note~1 and Supplementary Figure~1), which is more accurate than the tight binding model because it accounts for the waveguide amplitudes at different parts of each ring. As shown in Fig.~\ref{fig:floquet}a, a pair of edge states span the bandgap near the resonant frequency of individual site rings. The wavefunctions of this edge state pair are localized on opposite edges of the stripe (Figs.~\ref{fig:floquet}b,c).

The microring array discussed above exhibits a quantum spin Hall (QSH) phase for photons near a resonant frequency $\omega_0$ of the site ring. Since each site ring supports multiple resonances with different resonant frequencies separated by the free spectral range of the ring $\Omega\ll \omega_0$, if we take into account the frequency dimension, the system then corresponds to layers of decoupled 2D QSH state, each associated with a resonance of the site ring. The tight-binding Hamiltonian describing this system is
\begin{equation}
H^{3D}=\sum_m\left[\omega_m\sum_{x,y,\sigma}a^\dagger_{x,y,m,\sigma}a_{x,y,m,\sigma}-t\sum_{x,y,\sigma}\left(a^\dagger_{x+1,y,m,\sigma}a_{x,y,m,\sigma}+e^{i\sigma x\phi}a^\dagger_{x,y+1,m,\sigma}a_{x,y,m,\sigma}+h.c.\right)\right]
\label{eq:stacked-origin}
\end{equation}
where $a^\dagger_{x,y,m,\sigma}$ ($a_{x,y,m,\sigma}$) is the creation (annihilation) operator for the $m$-th order resonant mode of the site ring centered at $(x,y)$. To show explicitly that Eq.~\ref{eq:stacked-origin} is the same as stacking layers of 2D QSHE, we define $c_{x,y,m,\sigma}=a^{-i\omega_m t}a_{x,y,m,\sigma}$ and transform Eq.~\ref{eq:stacked-origin} to the rotating frame
\begin{equation}
H^{3D}=-t\sum_{m,x,y,\sigma}\left(c^\dagger_{x+1,y,m,\sigma}c_{x,y,m,\sigma}+e^{i\sigma x\phi}c^\dagger_{x,y+1,m,\sigma}c_{x,y,m,\sigma}+h.c.\right)
\label{eq:stacked}
\end{equation}
Eq.~\ref{eq:stacked} describes a 3D system labeled by $(x,y,m)$ consisting of stacked layers of 2D QSH with no interlayer coupling. The 3D system is gapped since the 2D magnetic bandgaps shown in Fig.~\ref{fig:floquet}a are preserved. The triad of weak topological indices of such a system are \cite{lu2016b,fu2007}
\begin{equation}
\nu_1=C^{yz}(k_x)=0,\;\nu_2=C^{zx}(k_y)=0,\;\nu_3=C^{xy}(k_z)=C^{2D}
\label{eq:chern}
\end{equation}
where $\nu_3=C^{xy}(k_z)$ is a first Chern number for the 3D topological insulator, and is defined as integral of Berry curvature $\mathcal{F}_{xy}(\mathbf{k})$ in the $k_x-k_y$ momentum plane within the first Brillouin zone. $\nu_1$ and $\nu_2$ are similarly defined. 

The result in Eq.~\ref{eq:chern} can be understood as follows. With no inter-layer coupling, the Fourier transform of the Hamiltonian in Eq.~\ref{eq:stacked} will be $k_z$-independent. Consequently the Berry curvature $\mathcal{F}_{yz}(\mathbf{k})=\mathcal{F}_{zx}(\mathbf{k})=0$ for all $\mathbf{k}$, and first two indices vanish. The third index $C^{xy}(k_z)$ is identical to the 2D QSH Chern number $C^{2D}$ in Eq.~\ref{eq:QSHE}. For an effective magnetic field of $+1/3$ ($-1/3$) flux quanta per unit cell for the $+$ ($-$) spin, $C^{2D}=\pm 1$ ($\mp 1$) for the lower and upper magnetic bandgap respectively \cite{thouless1982,osadchy}. Since the bulk spectrum is completely gapped, $C^{xy}(k_z)$ cannot change as a function of $k_z$. The same applies to the other two indices.

\subsection*{Screw dislocation and robust one-way frequency conversion}
To probe weak topological insulator, one needs to break crystal symmetry by cutting a surface or introducing a dislocation. In our system in the absence of dynamic modulation, there is a complete lack of interlayer coupling since the frequency does not change. Such a lack of coupling provides an ideal weak 3D TI with large topological bandgap in the bulk. However, it also means that any measurable surface topological signature is intrinsically 2D. To demonstrate a genuinely 3D signature of the topological phase, we propose to modulate a few rings in the array to introduce a single screw dislocation line into our 3D weak TI. It has been predicted that a single dislocation line in 3D TI can trap one-dimensional topologically protected modes \cite{ran2009,teo2010,lu2016b,bi2017}. For weak topological materials characterized by $\bG_\nu$ \cite{fu2007}, the number of 1D topologically protected modes trapped by a dislocation of Burger's vector $\bB$ is $\bG_\nu\cdot\bB/2\pi$. While it is technically challenging and energetically unfavorable \cite{kosterlitz1973} to create a single dislocation line in solid state materials, in our photonic platform with modulation a single dislocation can be controllably realized. In contrast to previous studies of screw dislocations which mostly focus on their signature in scattering of electrons in the far-field \cite{bird1988,bausch1998}
, here we are able to isolate and probe a single dislocation line and demonstrate its interaction with the topology of photonic bands. 

\begin{figure*}[hbt]
\includegraphics[width=6in]{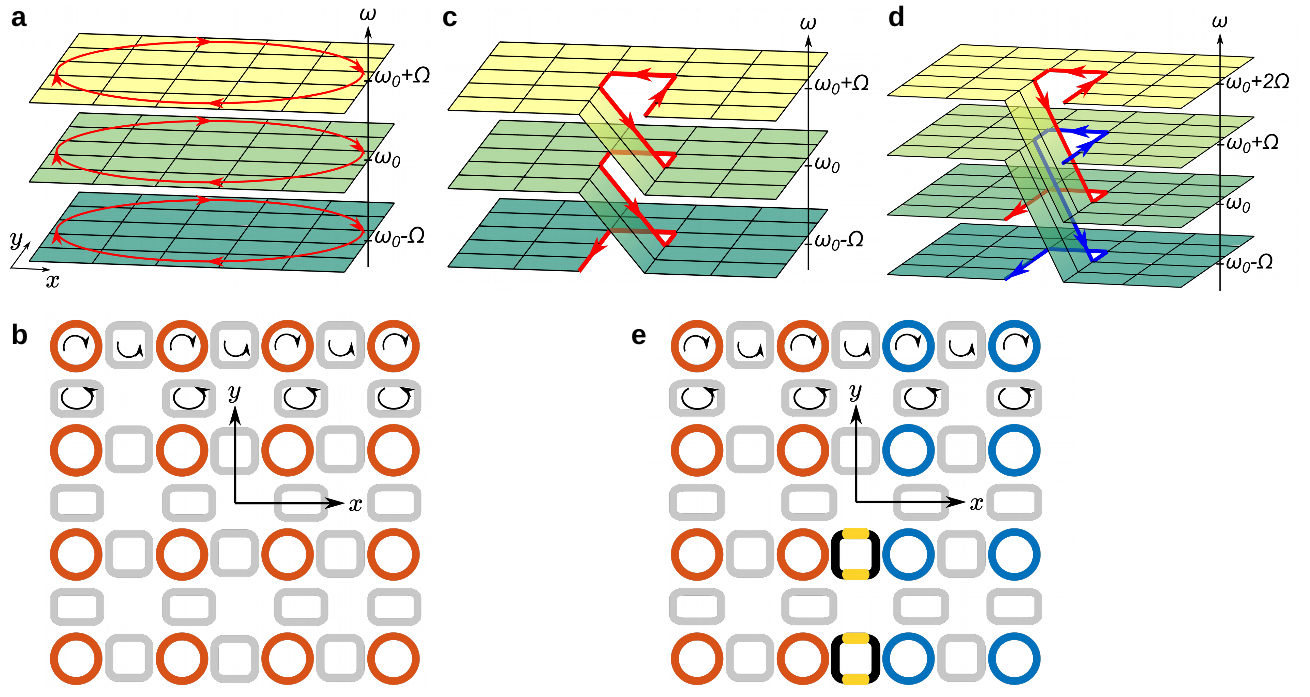}
\caption{3D weak topological insulator and screw dislocation. (a) A 3D lattice formed by stacking layers of 2D QSH states. The set of discrete resonances of a ring forms a lattice in the third, synthetic frequency dimension. (b) The microring array implementing (a). Red and grey rings are site and link rings with the same free spectral range $\Omega$. The link rings providing coupling along $y$ are spatially shifted to provide directional coupling phase that implements the Landau gauge. (c,d) Screw dislocations with Burger's vector $\bB=(0,0,-1)$ and $\bB=(0,0,-2)$, respectively. One-way topological state(s) spatially localized around the dislocation flows along the synthetic frequency dimension. (e) The microring array implementing (c,d). Red and blue rings are site rings with the same resonant frequencies but different waveguide dispersions. Grey rings are static link rings providing intra-layer couplings. Black rings are dynamic link rings whose index of refraction is modulated at a frequency equal to the free spectral range $\Omega$. It couples $\omega$ mode in the blue ring to $\omega+\Omega$ in the red ring, thus forming the inter-layer links shown in (c,d).}
\label{fig:setup}
\end{figure*}

\begin{figure*}[hbt]
\includegraphics[width=5.5in]{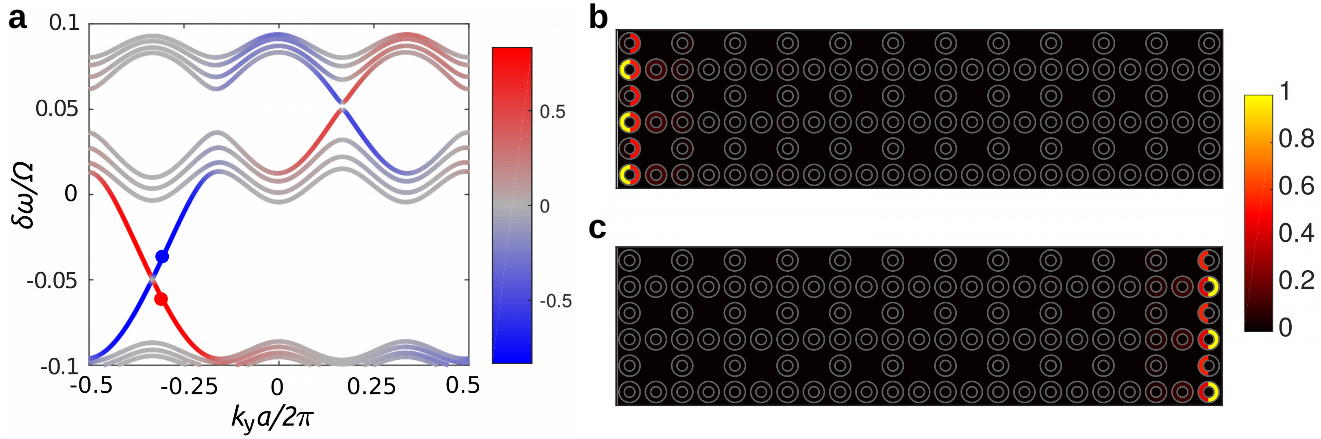}
\caption{2D QSHE in a microring array. (a) Band structure and (b,c) edge states of a two dimensional stripe of rings  with effective magnetic flux of $1/3$ quanta for the clockwise mode in the site ring. The strip is infinite along $y$ and has $12$ site rings along $x$. (a) Color scale for the band structure represents the difference of eigenstate modal intensity between the rings on the left and right edge of the structure. Thus the red and blue lines are edge states localized on the left and right edge of the stripe, respectively. The red and blue dots show the states plotted in (b) and (c), respectively. (b,c) Edge state intensity distribution, shown for three periods along $y$. The scattering amplitude at each site ring to link ring coupler is $\gamma_s=0.75$.}
\label{fig:floquet}
\end{figure*}

To introduce a screw dislocation as shown in Fig.~\ref{fig:setup}c, we replace a few in-plane coupling terms in Eq.~\ref{eq:stacked} with inter-layer coupling:
\begin{align*}
c^\dagger_{x+1,y,m,\sigma}c_{x,y,m,\sigma}+h.c.\to c^\dagger_{x+1,y,m-1,\sigma}c_{x,y,m,\sigma}+h.c.,\;\;\forall x=-0.5,\;y<0
\end{align*}
This introduces a branch cut at $x=0,\;y<0$. Away from the cut, the Hamiltonian is unchanged from Eq.~\ref{eq:stacked}. Across the branch cut, tunneling towards $+x$ ($-x$) is associated with moving down (up) one layer along the frequency dimension, which results in a screw dislocation located at $(0,0)$. 

A microring array that implements such a screw dislocation in the above mentioned 3D TI is shown in Fig.~\ref{fig:setup}e. The structure is similar to Fig.~\ref{fig:setup}b which exhibits the 2D QSH phase. In particular, all the rings have the same free spectral range $\Omega$, and all the site rings have the same resonant frequency, which coincides with the anti-resonance of the static link rings. The differences from the 2D QSH configuration are: (1) The waveguide forming the site rings at $x>0$ (blue) and $x<0$ (red) have different wavevectors for the same resonant frequency. This can be implemented with a different refractive index or cross-sectional dimension of the waveguide. (2) We replace the static link rings (grey) on the branch cut with dynamic link rings (black). The details of one of these modified link rings are provided in Fig.~\ref{fig:device}a, and in Supplementary Figures~2-4. It consists of a slot waveguide that supports an even and an odd mode. A direct photonic transition between these two modes is induced by dynamic modulation of the refractive index at the frequency $\Omega$ that is equal to the free spectral range of the site rings \cite{fang2012a,tzuang2014}. Outside of the modulation region, the slot waveguide gap is tapered down so that the wavevectors of the even and odd mode differ. The odd mode is phase-matched to the red site ring. while the even mode is phase-matched to the blue site ring. The phase-mismatched couplings are minimized. Consequently, the dynamic link ring couples the $m$-th resonant modes in the blue ring to the $m+1$-th resonant modes in the red ring (Fig.~\ref{fig:device}b). Hopping counter-clockwisely around a closed path encircling the center of the array in Fig.~\ref{fig:setup}e decreases the frequency by $\Omega$. Thus a screw dislocation with Burgers vector $\bB=(0,0,-1)$ is created at the center of the array. Practical design considerations are discussed in Supplementary Notes 2\&3, and an example is provided in Supplementary Table~1.

We consider a $12\times 12$ array of microrings with a screw dislocation at the center. We assume modes extend infinitely along the frequency axis, and use periodic boundary conditions along $x$ and $y$. The resulting spectrum as a function of the wavevector $k_f$ along the dislocation line is plotted in Fig.~\ref{fig:bs}a. The dislocation line at $(0,0)$ indeed supports a mode that spans the topological bandgap and propagates unidirectionally down the frequency axis, as shown by the red line in Fig.~\ref{fig:bs}a. The corresponding eigenstate wavefunction is well localized on the rings immediately surrounding the dislocation line at the center of the array （Fig.~\ref{fig:bs}b）, and is minimally affected by the finite size effect at the edges of the array. As an artifact of the periodic boundary condition, a second dislocation line with an opposite Burgers vector is present at the boundary of the array. It supports a one-way state in the topological bandgap propagating in the opposite direction, as shown by the blue line in Fig.~\ref{fig:bs}a, with an eigenstate localized around the dislocation center at $(0,6)$ (Fig.~\ref{fig:bs}c).

Multimode one-way photonic waveguides can increase the capacity of the unidirectional guiding channel and enable new devices and functionalities \cite{Scott2014,Wang15}. In our system, the number of topological modes supported by the dislocation is $\bG_\nu \cdot\bB/2\pi$ \cite{ran2009,lu2016b,bi2017}. Multimode one-way waveguide can be realized by increasing either the amplitude of dislocation or the Chern number. We can double the value of $\bB$ by modulating the dynamic link ring at frequency $2\Omega$. Each dynamic link ring connects states that are two resonances apart between the neighboring site rings, and we effectively have two interleaved copies of the spiral surface shown in Fig.~\ref{fig:setup}d, each supporting a 1D topological state at the dislocation line. This is confirmed by the band diagram shown in Fig.~\ref{fig:bs}c where two pairs of one-way states span the topological gap. The wavefunction projected onto the $x-y$ plane is similar to Fig.~\ref{fig:bs}b for both pairs. Alternatively, one can access magnetic bandgaps with a higher Chern number by changing the magnetic flux \cite{thouless1982,osadchy}. In this case, the multiple gap states supported at the dislocation will have orthogonal spatial wavefunctions and similar band dispersions. 

The dislocation state can be experimentally probed by coupling a single frequency signal within the magnetic bandgap into one of the site rings around the dislocation at $(0,0)$ through an external waveguide. Fig.~\ref{fig:input} plots the steady state power distribution in the presence of realistic waveguide losses when such a continuous wave input at frequency detuning $\delta\omega=-0.03\Omega$ from a resonance is coupled to the site ring at $(-0.5,0.5)$. As shown in Fig.~\ref{fig:input}a, the input at sideband $m=0$ is unidirectionally transported to lower frequency resonances. The structure thus enables one-way frequency conversion. In our simulation, a $5\%$ round-trip power loss is assumed for each ring, thus the total power decays while traveling down the frequency axis. To test the robustness of the one-way frequency conversion, we introduce random fluctuations into the resonant frequency and the inter-ring coupling strength. We first introduce a random detuning of the resonance frequency $\omega_0$ for each ring with the distribution $\Delta\omega_{0,i}/\Omega\sim \mathcal{N}(0,0.05)$, where $i$ labels individual rings in the array. Fig.~\ref{fig:input}b shows the mean (red dots) and standard deviation (grey area) of the spectral power distribution averaged over $20$ randomly generated configurations. The unidirectionality of the frequency conversion is well preserved. We then introduce a random variation into the inter-ring coupling strength $\gamma_0$ for each pair of adjacent rings such that $\Delta\gamma_{\langle ij\rangle}/\gamma_0\sim\mathcal{N}(0,0.1)$ for each adjacent pair of rings. $\gamma_{\langle ij\rangle}=\Delta\gamma_{\langle ij\rangle}+\gamma_0$ is capped to be within $[0,1]$. The result shown in Fig.~\ref{fig:input}c again demonstrates the robustness of the one-way frequency conversion. Finally we examine the spatial power distribution at different sidebands. As shown in Fig.~\ref{fig:input}d, the power is concentrated on the rings immediately surrounding the dislocation line, and propagates counter-clockwisely around it. In our simulation, we imposed a hard frequency boundary condition at $m=\pm 5$ similar to that shown in Fig.~\ref{fig:setup}a. When the signal reaches the bottom frequency layer at $m=-5$, it travels along the branch cut defined by the dynamic link rings and merges into a clockwise edge state at the boundary of the array (Fig.~\ref{fig:input}d). 

In summary, we show that existing implementation of 2D topological photonic states based on 2D microring arrays can be extended to implement 3D weak topological insulators. This allows us to study 3D topological effects using on-chip integrated photonic platform, bypassing the difficulty of constructing complex 3D optical structures. Our proposal also yields a sizable 3D topological bandgap comparable to its 2D counterpart, which provides a clean platform to study the interaction of crystallographic defect with band topology. As an example, our system can be used to demonstrate the robust unidirectional 1D waveguide pinned by a screw dislocation in a 3D TI, implemented through dynamic modulation of a few microrings. 


\section*{Acknowledgements}
This work is supported by U.S. Air Force of Scientific Research Grant No.FA9550-17-1-0002 and U.S. National Science Foundation Grant No.CBET-1641069. Q.L. acknowledges the support of a Stanford Graduate Fellowship. X.-Q.S. and S.-C.Z. acknowledge support from the US Department of Energy, Office of Basic Energy Sciences under contract DE-AC02-76SF00515.

\section*{Author contributions}
Q.~L. and S.~F. conceived the idea of constructing a three-dimension photonic topological insulator in a ring resonator lattice. Q.~L., X.-Q.~S. and M.~X. conceived the scheme for realizing a screw dislocation. All authors contributed to the discussion of the results and writing of the manuscript. S. F. supervised the project.

\section*{COMPETING FINANCIAL INTERESTS}
The authors declare no competing financial interests.

\pagebreak

\begin{figure*}[hbt]
\includegraphics[width=5in]{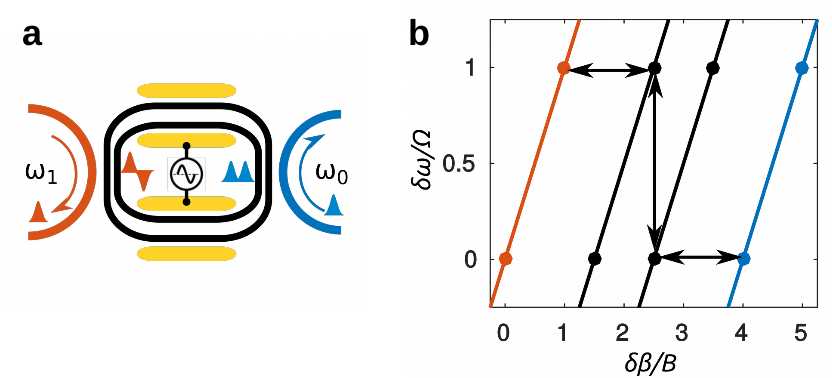}
\caption{Design of the dynamic link ring. (a) A dynamic link ring formed by a slot waveguide. It supports two modes with even and odd transverse profiles, respectively. The yellow pads are radio-frequency (RF) modulators that dynamically vary the refractive index of the waveguide. This drives a direct photonic transition between the even and odd modes. The width of the air gap in the slot waveguide at the modulated region is chosen such that the frequency separation between the even and odd modes matches the modulation frequency. At the waveguide coupler to the site rings, this air gap is tapered down to provide phase-matched coupling to the site rings. (b) Waveguide dispersion and ring resonance of the site rings and the dynamic link ring. $\Omega$ is the free spectral range of the rings, and $B$ is the wavevector difference between adjacent resonances in each ring. The red line corresponds to the dispersion of waveguide mode in the red site ring as well as the odd mode in the dynamic link ring at the waveguide coupler region. The blue line corresponds to the dispersion of the waveguide mode in the blue site ring as well as the even mode in the dynamic link ring at the waveguide coupler region. The black lines correspond to the dispersion of the even and odd waveguide modes of the dynamic link ring at the modulator region. Dots indicate the resonances of each ring. Black arrows show the coupling of a lower frequency mode in the blue site ring to a higher frequency mode in the red site ring through a driven transition in the dynamic link ring.}
\label{fig:device}
\end{figure*}

\begin{figure*}
\includegraphics[width=6.5in]{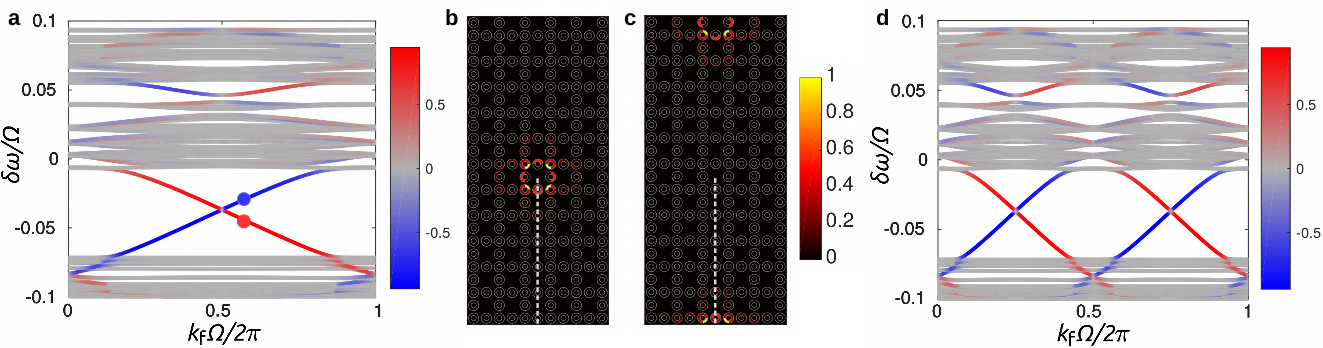}
\caption{1D topological state trapped by a screw dislocation. Calculated for the lattice with a screw dislocation shown in Fig.~1, using an $12\times12$ array of site rings. Assume infinite resonant modes along the synthetic frequency dimension, and periodic boundary conditions in both $x$ and $y$. (a,d) Band diagram for a lattice with $\bB=(0,0,-1)$ and $\bB=(0,0,-2)$, respectively. Color scale represents the difference of eigenstate modal intensity between rings surrounding the dislocation lines centered at $(0,0)$ and $(0,6)$. Red lines are one-way states localized on the dislocation centered at $(0,0)$. Blue lines are one-way states localized on the dislocation centered at $(0,6)$, as a result of the periodic boundary condition. Red and blue dots are the states plotted in (b) and (c). (b,c) Eigenstate intensity distribution. For simplicity, only $6$ columns closest to the center are shown. The dashed line indicates the location of dynamic link rings. For the dynamic link ring, the intensity of the even and odd modes are simply added together for display.}
\label{fig:bs}
\end{figure*}

\begin{figure*}
\includegraphics[width=6in]{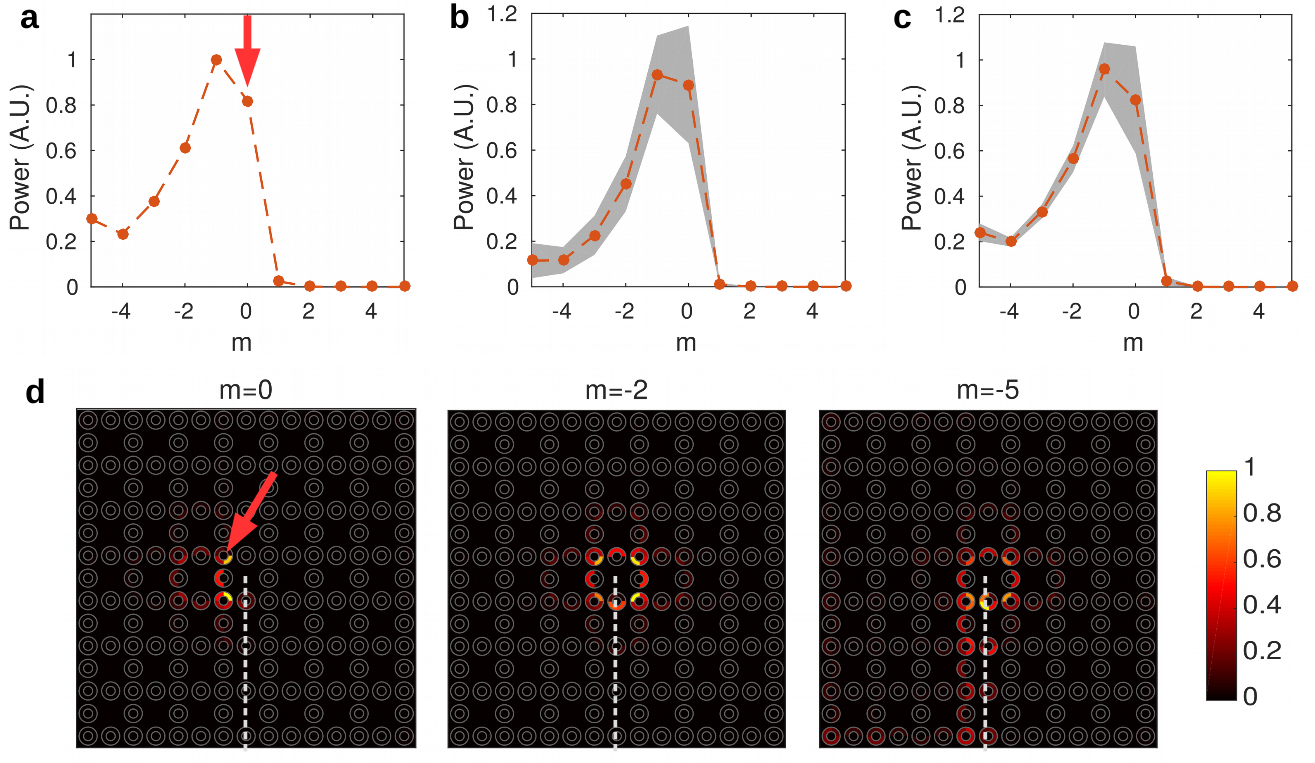}
\caption{Probing the dislocation state. A screw dislocation is created in a $8\times 8$ array of site rings. A single frequency input is coupled into a site ring at the center through an external waveguide. $11$ sidebands are used, with hard truncation along the frequency boundary \cite{yuan2016}. $m$ is the order of sidebands, where $m=0$ is the input frequency. The input frequency is in the lower magnetic bandgap with a detuning $\delta\omega=-0.03\Omega$ from the site ring resonance. The red arrows in (a,d) indicate the  input resonant mode and spatial location of the input ring. (a) Steady state aggregated intensity at each sideband. One-way frequency down conversion is observed. (b) Adding a normal distributed noise to the resonance of all rings with standard deviation $\sigma=0.05$ for $\delta\omega/\Omega$. Red dots show the mean intensity, and grey areas the standard deviation. Averaged over $20$ configurations. (c) Adding a normal distributed noise to the coupling strength between all rings with standard deviation $\sigma=0.1$ for the ratio to the original value. Red dots show the mean intensity, and grey areas the standard deviation. Averaged over $20$ configurations. (d) Spatial intensity distribution at three different sidebands. Power is largely localized at the rings immediately surrounding the dislocation line. When the power reaches the bottom edge ($m=-5$), it continues along the branch cut and merges into an edge state of the stripe. }
\label{fig:input}
\end{figure*}
\end{document}